\documentclass[journal=jacsat,manuscript=communication]{achemso}

\newcommand*{\ie}{\emph{i.\,e.,}\ }

\newcommand*{\dEscdE}{{\rm d}E_{\rm sc}/{\rm d}E}
\newcommand*{\dEhbdE}{{\rm d}E_{\rm hb}/{\rm d}E}

\author{Tristan Bereau}
\affiliation[Carnegie Mellon University]
{Dept. of Physics, Carnegie Mellon University, Pittsburgh, PA 15213, USA}
\author{Michael Bachmann}
\affiliation[Forschungszentrum J\"ulich]
{Institut f\"ur Festk\"orperforschung, Theorie II, Forschungszentrum J\"ulich,
  52425, J\"ulich, Germany}
\author{Markus Deserno}
\email{deserno@andrew.cmu.edu}
\affiliation[Carnegie Mellon University]
{Dept. of Physics, Carnegie Mellon University, Pittsburgh, PA 15213, USA}

\title[Protein folding cooperativity] 
{Interplay between secondary and tertiary structure formation \\in protein
  folding cooperativity}

\begin{document}
\begin{abstract}
  Protein folding cooperativity is defined by the nature of the
  finite-size thermodynamic transition exhibited upon folding:
  two-state transitions show a free energy barrier between the folded
  and unfolded ensembles, while downhill folding is barrierless. A
  microcanonical analysis, where the energy is the natural variable,
  has shown better suited to unambiguously characterize the nature of
  the transition compared to its canonical counterpart. Replica
  exchange molecular dynamics simulations of a high resolution
  coarse-grained model allow for the accurate evaluation of the
  density of states, in order to extract precise thermodynamic
  information, and measure its impact on structural features. The
  method is applied to three helical peptides: a short helix shows
  sharp features of a two-state folder, while a longer helix and a
  three-helix bundle exhibit downhill and two-state transitions,
  respectively. Extending the results of lattice simulations and
  theoretical models, we find that it is the interplay between
  secondary structure and the loss of non-native tertiary contacts
  which determines the nature of the transition. 
\end{abstract}

The folding cooperativity of proteins is characterized by the relative
population of intermediate states at the transition temperature: while
two-state transitions exhibit two energetic peaks characterizing the folded
and unfolded ensembles, downhill folders show a unimodal distribution of
energetic states without any barrier\cite{privalov79, privalov82}. This aspect
of finite-size thermodynamic transitions can provide insight into the folding
mechanism, but energetic populations can be difficult to measure. Therefore,
protein folding cooperativity is often probed using the calorimetric criterion
\cite{calo_chan}, which quantifies the sharpness of the specific heat
curve. From a computer simulation point of view, however, evaluating the
probability density $p(E)$ remains an appealing idea, as it would provide an
unambiguous description of the thermodynamic transition. Although currently
untractable atomistically due to sampling limitations, high resolution
coarse-grained models offer an alternative approach\cite{voth}. While cutting
down significantly on computational time, they can retain much chemical
detail, and some are even able to fold simple peptides with no prior knowledge
of the native state. In this Communication we study the link between
thermodynamics and structure for helical peptides using such a coarse-grained
model,\cite{bereaudeserno09} details of which can be found in the Supporting
Information.

To characterize the thermodynamics of finite-size systems, it has been
shown that a microcanonical analysis, based on the
entropy $S(E)$, is often more informative than a canonical
analysis\cite{gross, huller}. Microcanonically, $S(E)=k_{\rm
  B}\ln\Omega(E)$ where $\Omega(E)$ is the density of states. One
remarkable feature of such a description is its ability
to unambiguously distinguish between discontinuous (\ie two-state) and
continuous (\ie downhill) transitions. Indeed,
two-state transitions exhibit a depletion of intermediate energetic
states leading to local convexity in the entropy. This can be best
observed by defining the quantity $\Delta S(E)=\mathcal{H}(E)-S(E)$,
where the first term is the (double-)tangent to $S(E)$
in the transition region\cite{deserno97, jungbach06, jungbach08}. The
method relies on accurate measurements of the density of states,
calculated here using the Weighted Histogram Analysis
Method\cite{kumar95}. All order parameters will be analyzed as a
function of energy.

\begin{figure}[t]
  \includegraphics[width=\linewidth]{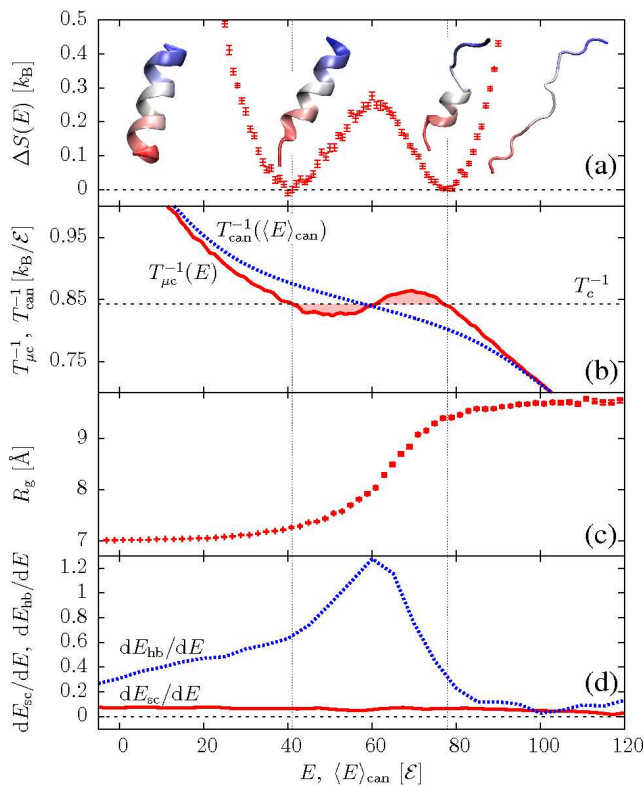}
  \caption{(\texttt{AAQAA})$_3$. (a) $\Delta S(E)$; error bars reflect the
    variance of the data points ($1\,\sigma$ interval). (b) inverse
    temperature from a canonical $T^{-1}_{\rm can}(\langle E\rangle_{\rm
      can})$ and microcanonical $T^{-1}_{\mu \rm c}(E)=\partial S/\partial E$
    analysis, where $\langle E\rangle_{\rm can}$ is the canonical average
    energy. (c) radius of gyration $R_{\rm g}(E)$ with the error of the
    mean. (d) rates of H-bond and side chain energies $\dEhbdE,\
    \dEscdE$. Vertical lines delimit the transition region; its width
    corresponds to the microcanonical latent heat $\Delta Q$.}
  \label{fig:n3}
\end{figure}

We first examine a short $\alpha$-helix of sequence
(\texttt{AAQAA})$_3$\cite{aaqaa3_exp}. The density of states reveals a
discontinuous transition with nonzero latent heat $\Delta Q$ (\ref{fig:n3},
a) between the folded and unfolded ensembles (representative conformations at
different energies are shown). Phase coexistence is associated with a
backbending\cite{gross} of the microcanonical inverse temperature $T^{-1}_{\mu
  \rm c}=\partial S/\partial E$ while the corresponding canonical relation
$T^{-1}(\langle E\rangle_{\rm can})$, where $\langle E\rangle_{\rm can}$ is
the average energy, is monotonic (\ref{fig:n3}, b). The radius of gyration
(\ref{fig:n3}, c) quickly drops inside the coexistence region, indicating that
most structural rearrangements happen within this energy interval. We will
assume the hydrogen-bond and side chain energies, $E_{\rm hb}$ and $E_{\rm
  sc}$, to be suitable proxies of secondary structure and tertiary contacts,
repectively. It proves instructive to look at their energetic rates, $\dEhbdE$
and $\dEscdE$ (\ref{fig:n3}, d): even though $\dEscdE$ stays virtually flat
over the energy range considered, the sharp peak in $\dEhbdE$ indicates that
most secondary structure forms within the coexistence region.

\begin{figure}[t]
  \includegraphics[width=\linewidth]{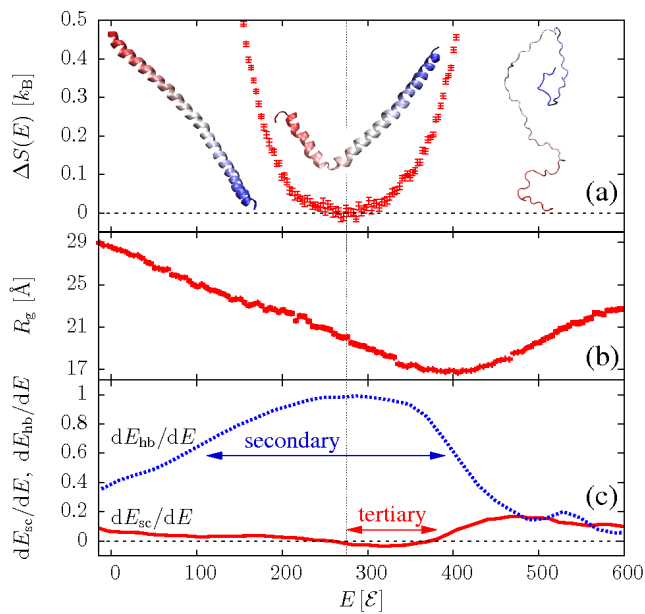}
  \caption{(\texttt{AAQAA})$_{15}$. (a) $\Delta
    S(E)$. (b) radius of gyration $R_{\rm g}(E)$. (c) rates of H-bond
    and side chain energies $\dEhbdE,\ \dEscdE$. Horizontal arrows
    indicate where most secondary structure forms and where non-native
    tertiary contacts dissolve. The vertical line marks the transition
    point.}
  \label{fig:n15}
\end{figure}

Elongating the sequence to (\texttt{AAQAA})$_{15}$ leads to a qualitative
change in the folding mechanism. The ground state again forms a single
$\alpha$-helix, but the transition is now continuous: as can be seen in
\ref{fig:n15} (a) and Figure S2 (Supporting Information), there is a single
transition point and the latent heat is zero. The radius of gyration
(\ref{fig:n15}, b) features a sharp minimum \emph{above} the transition point,
indicative of a chain collapse into ``maximally compact non-native
states.''\cite{dillstigter95} Upon lowering the energy further, the chain will
reorganize from such non-native states into the helical state. In doing so,
the rate of tertiary contact formation $\dEscdE$ dips below zero
(\ref{fig:n15}, c), hence there is an energetic penalty associated with
tertiary rearrangements.  Hydrogen-bond formation occurs over a large
energetic interval as indicated by the broad maximum in $\dEhbdE$.  The
absence of any two-state signal is consistent with theoretical models of the
helix-coil transition\cite{zimmbragg}: the energetic cost of breaking a
hydrogen-bond is outweighed by the conformational entropy gained. Further
analysis indicates on average two helices at the transition point.

\begin{figure}[t]
  \includegraphics[width=\linewidth]{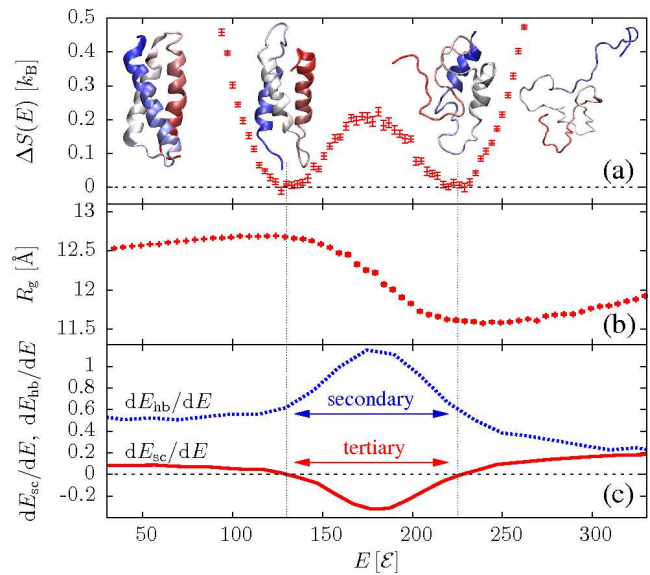}
  \caption{Three-helix bundle $\alpha$3D. (a) $\Delta S(E)$. (b)
    radius of gyration $R_{\rm g}(E)$. (c) rates of H-bond and side
    chain energies $\dEhbdE,\ \dEscdE$.}
  \label{fig:a3d}
\end{figure}

While of similar length, the 73 amino acid \emph{de novo} three-helix bundle
$\alpha$3D (PDB code: 2A3D)\cite{a3d_exp} does show a discontinuous
transition, see \ref{fig:a3d} (a). Representative conformations sampled in the
two coexisting ensembles stand as good proxies of the ground state and
unfolded state, unlike for the downhill folding transition of
(\texttt{AAQAA})$_{15}$. The radius of gyration again shows a minimum above
the transition (\ref{fig:a3d}, b), and folding once more starts from maximally
compact non-native states. Notice that secondary structure formation and the
loss of non-native tertiary contacts (\ref{fig:a3d}, c) are sharp and
predominantly localized within the coexistence region. The three helices form
inside the same energetic interval due to the inter-helical cooperativity
imprinted in the sequence\cite{ghoshdill09}. Chain compaction is
  due to strong side chain-side chain interactions.

Overall, we can correlate thermodynamic features with structural information
from the three peptides studied here. While (\texttt{AAQAA})$_3$ is too short
for tertiary interactions to play any role, the transitions associated with
(\texttt{AAQAA})$_{15}$ and the bundle $\alpha$3D are both associated with
tertiary rearrangements. These two examples suggest that independently of its
nature, the folding transition is driven by the loss of non-native tertiary
contacts (\ie the region where $\dEscdE < 0$)---reminiscent of the
heteropolymer collapse model\cite{dillstigter95}. On the other hand, secondary
structure formation has shown very different signals: (\texttt{AAQAA})$_3$ and
$\alpha$3D exhibit sharp peaks where (\texttt{AAQAA})$_{15}$ displays a broad
maximum.  As shown in \ref{fig:n15} (c), secondary structure formation in a
downhill folding peptide occurs over a much broader interval compared to the
loss of non-native tertiary contacts, whereas these two quantities are
contained within the same narrow interval for a two-state peptide
(\ref{fig:a3d}, c).  Cooperative secondary and tertiary structure formation
has been proposed as a mechanism for two-state folding from lattice
simulations\cite{kayachan00} and theoretical models\cite{ghoshdill09}. Beyond
this, our results also highlight the interplay between secondary structure
\emph{formation} and the \emph{loss} of non-native tertiary contacts. Our
conclusions on the thermodynamics of the short-, long-, and bundled helix are
compatible with the calorimetric criterion, which gives, respectively,
$\delta=0.78,\ 0.52,\ 0.78$ for the calorimetric ratio\cite{calo_chan}. Note
that Ghosh and Dill\cite{ghoshdill09} predicted $\delta=0.72$ for the similar
bundle $\alpha$3C. However, the main strength of a microcanonical analysis
stems from an access to fine aspects of thermodynamic information that are
otherwise difficult to obtain either canonically or from experiments. It thus
stands as a complementary tool to gain further insight.

\acknowledgement We acknowledge stimulating discussions with K. Binder,
W. Paul, R.H. Swendsen, and M. Taylor, and funding through the NIH grant
P01AG03231. MB thanks the Forschungszentrum J\"ulich for supercomputer time
grants jiff39 and jiff43. TB acknowledges support from the
  Astrid and Bruce McWilliams Fellowship.

\suppinfo

Peptide model; simulation and analysis methods.


\end{document}